\documentclass[pdflatex,sn-mathphys-num]{sn-jnl}
\usepackage[utf8]{inputenc}
\usepackage[T1]{fontenc}
\usepackage{graphicx}%
\usepackage{multirow}%
\usepackage{amsmath,amssymb,amsfonts}%
\usepackage{amsthm}%
\usepackage{mathrsfs}%
\usepackage[title]{appendix}%
\usepackage{xcolor}%
\usepackage{textcomp}%
\usepackage{manyfoot}%
\usepackage{booktabs}%
\usepackage{algorithm}%
\usepackage{algorithmicx}%
\usepackage{algpseudocode}%
\usepackage{listings}%
\usepackage{makecell}
\usepackage{latexsym}
\usepackage{array}
\usepackage{tabularx}
\usepackage{multirow}
\usepackage{amsfonts}
\usepackage{pifont}
\usepackage{array}
\usepackage{tikz}
\usetikzlibrary{tikzmark,calc}
\usepackage{colortbl}
\usepackage{xcolor}
\usepackage{pst-all} 
\usepackage{pst-poly}
\newpsobject{showgrid}{psgrid}{subgriddiv=1,griddots=10,gridlabels=6pt}
\usepackage{graphicx}
\usepackage{fancybox}
\usepackage{wrapfig}
\usepackage{todonotes}
\graphicspath{{./}{./fig/}}

\usepackage{caption}
\theoremstyle{thmstyleone}%
\theoremstyle{thmstyletwo}%
\theoremstyle{thmstylethree}%

\begin{document}

\title{Explainable Verification of Hierarchical Workflows Mined from Event Logs with Shapley Values}


\author{\centering Rados{\l}aw Klimek \& Jakub Błażowski\\
AGH University of Krakow\\
rklimek@agh.edu.pl}



\abstract{Workflow mining discovers hierarchical process trees from event logs,
but it remains unclear why such models satisfy or violate logical properties,
or how individual elements contribute to overall behavior.
We propose to translate mined workflows into logical specifications and analyze
properties such as satisfiability, liveness, and safety with automated theorem provers.
On this basis, we adapt Shapley values from cooperative game theory to attribute outcomes
to workflow elements and quantify their contributions.
Experiments on benchmark datasets show that this combination identifies critical nodes,
reveals redundancies, and exposes harmful structures.
This outlines a novel direction for explainable workflow analysis with direct relevance to software engineering practice,
supporting compliance checks, process optimization, redundancy reduction, and the design of next-generation process mining tools.}

\maketitle

\section{Introduction}

Workflow mining is a cornerstone of process analysis: algorithms such as inductive mining efficiently discover hierarchical process trees from large event logs.
These models capture real behavior with operators like sequence, choice, parallelism, and loops, and are used in business optimization and software process monitoring.
Yet, as workflows grow in size and complexity, it becomes difficult for software engineers and
process analysts to understand why a model satisfies or violates certain properties, or how individual elements contribute.
Existing approaches provide global property results or conformance measures, but offer little insight into the role of specific workflow nodes.

We propose to translate workflows into logical specifications and analyze satisfiability, liveness, and safety with automated theorem provers.
On top of this, we adapt Shapley values from cooperative game theory to measure the contribution of each element, yielding rankings of critical, redundant, and even harmful nodes.
This extends property analysis with explainability, turning black-box results into interpretable structures.

Experiments on benchmark datasets (BPI Challenge 2012, Hospital Billing, Running Example) show that the approach exposes subtle structures, highlights essential nodes, and reveals redundancies.
These findings outline an emerging direction toward explainable workflow analysis, combining the rigor of formal methods with the interpretability of game-theoretic attribution,
laying a foundation for next-generation tools in software analytics and compliance verification.
For software engineers, these explanations enable compliance checks, reduce redundancy,
and target process optimizations in large-scale workflows.
They help identify fragile fragments, support debugging of process-oriented systems, and make analysis outcomes understandable to stakeholders.

Prior work in process mining has emphasized conformance checking and alignments~\cite{Rozinat-vanderAalst-2008,van-der-Aalst-2016-ProcessMining},
or explainability in machine learning (e.g., SHAP, LIME~\cite{Lundberg-Lee-2017,Ribeiro-etal-2016-LIME}), but none combine workflow property analysis with cooperative game-theoretic attribution.
This paper introduces such a combination as a foundation for explainable workflow analysis.
To our knowledge, no prior work combines logical property analysis of workflows
with cooperative game-theoretic attribution.
This novelty positions our approach as a first step toward explainable workflow analysis,
bridging process mining from event logs, automated reasoning with theorem provers rooted in formal logic, and explainable AI,
grounded in cooperative game theory and carried out in a logical style,
to make workflow analysis both qualitative and quantitative, as well as rigorous and interpretable.

This work is guided by three \textbf{research questions}:
(RQ1)~how logical workflow analysis can be enriched with explanations that attribute property satisfaction or violation to specific elements,
(RQ2)~whether Shapley values provide meaningful and stable rankings of critical, redundant, and harmful nodes,
and (RQ3)~what trade-offs arise between approximation strategies (Monte Carlo permutations vs.\ random subset sampling) in terms of scalability, stability, and interpretability.
Our \textbf{contributions} are as follows.
(i)~We adapt Shapley values from cooperative game theory to quantify the marginal contribution of workflow elements to logical properties.
(ii)~We evaluate two approximation strategies that make Shapley-based attribution feasible for non-trivial workflows.
(iii)~Finally, we provide preliminary evidence on benchmark datasets that the approach can support model diagnosis and simplification.

\section{Logical Verification and Shapley Attribution}

\subsection{Workflows as Logical Specifications}
Hierarchical workflows are discovered from event logs as process trees,
with internal nodes capturing control-flow operators such as sequence,
parallelism, choice, and loops, and leaves denoting activities.
To enable formal reasoning, each tree is systematically translated into
an equivalent logical specification.

The translation is pattern-based: a fixed set of behavioral patterns
(Seq, Xor, And, Loop) suffices to express process tree structures.
Each pattern has an associated schema of propositional temporal logic
formulas, ensuring preservation of properties such as satisfiability and
liveness. By composing these predefined fragments according to the tree,
a complete logical specification is obtained automatically~\cite{Klimek-Witek-2024-ASE-RENE}.

This encoding bridges process mining and formal verification.
The resulting specification is machine-checkable and can be analyzed with
existing theorem provers such as Vampire~\cite{Vampire-tool,Riazanov-Voronkov-2002}
and E~\cite{E-tool,Schulz-2002}.
The main properties of interest are satisfiability (internal consistency),
liveness (progress can always occur), and safety (undesired states excluded).
Because the mapping is automatic and compositional, it scales to large
workflows and provides the foundation for the attribution analyses
introduced in this paper.

\subsection{Players, Coalitions and Shapley Values}

The starting point of our analysis is a hierarchical process tree obtained by workflow mining,
typically with the inductive miner~\cite{Leemans-etal-2013,Augusto-etal-2019},
which constructs sound and block-structured trees even from large logs.
Such a tree serves as the basis for generating a logical specification,
which in turn enables attribution of property outcomes using Shapley values.

Each workflow element (tree node) is treated as a \emph{player} in a cooperative game.
A coalition is a subset of nodes retained in the model, with missing ones replaced by a neutral $\tau$ element.
The value function $v(C)$ evaluates the logical specification induced by coalition $C$:
\vspace{-0.3em}
\begin{itemize}
  \item $v(C)=1$ if the specification is satisfiable,
  \item $v(C)=1$ if liveness holds,
  \item $v(C)=1$ if safety constraints are respected,
\end{itemize}
and $v(C)=0$ otherwise.

Satisfiability is a prerequisite for meaningful reasoning, while
\emph{liveness} and \emph{safety} together provide a complete view of behavior~\cite{Kindler-1994,Alpern-Schneider-1985,Manna-Pnueli-1992}.

The Shapley value $\phi_i$ of a node $i$ is defined as
\[
\phi_i = \sum_{S \subseteq N \setminus \{i\}}
\frac{|S|!(n-|S|-1)!}{n!} \cdot [v(S \cup \{i\}) - v(S)],
\]
where $N$ is the set of all nodes and $S$ a coalition excluding $i$.
It captures the marginal contribution of element $i$ to property satisfaction,
linking local workflow structure with global outcomes.

%

\subsection{Approximation Strategies}
Exact Shapley computation requires $2^n$ coalitions, which is infeasible for realistic workflows.
We therefore use two approximation strategies:

\begin{enumerate}
  \item \textbf{Monte Carlo permutations:} random player orderings are sampled, players are added in turn, and marginal contributions recorded.
  This method is unbiased and faithful to the definition, but computationally expensive.
  \item \textbf{Random subset sampling:} subsets with and without a node are compared to estimate its effect.
  This scales better but introduces bias and higher variance.
\end{enumerate}

In practice, Monte Carlo provides more stable estimates, whereas random subsets trade accuracy for efficiency.

\section{Preliminary Results}

\subsection{Datasets and Setup}
\label{sec:Datasets-Setup}

We evaluate the approach on three benchmark datasets:
\emph{BPI Challenge 2012}\footnote{Available at \url{https://data.4tu.nl/articles/_/12689204/1} (last accessed on 20.07.2025)},
\emph{Hospital Billing}\footnote{Available at \url{https://data.4tu.nl/articles/_/12705113/1} (last accessed on 20.07.2025)}, and
the \emph{Running Example}\footnote{Available at \url{https://github.com/process-intelligence-solutions/pm4py/blob/release/tests/input_data/running-example.xes} (last accessed on 20.07.2025)}.
The first two represent real-world processes of different complexity,
while the latter is a small synthetic model often used for teaching.
To test robustness, workflows were mined at four noise thresholds (0.0, 0.25, 0.5, 1.0).
Here, \emph{noise} denotes infrequent or spurious activities in event logs,
and the threshold parameter controls how much such behavior is filtered out during discovery.
Logical properties of satisfiability, liveness, and safety were checked with
the Vampire~\cite{Vampire-tool} and E~\cite{E-tool} provers.

A \emph{configuration} is a dataset–noise–property triple, giving 36 in total.
For each configuration we verified coalitions of workflow elements, with caching used to avoid redundant checks.
All properties were tested in TPTP~\cite{Sutcliffe-2017}
format with a 2-second timeout.
Vampire proved more efficient, so most results rely on it.
In total, about 756,000 queries were executed ($\approx$21,000 per configuration),
with caching reducing repeats to $\sim$2,000.

Each coalition was evaluated against three properties.
\emph{Satisfiability} required \texttt{SZS = Satisfiable}~\cite{Schreiner-2023,Sutcliffe-2017},
ensuring the specification is consistent.
\emph{Liveness} was checked by $\Diamond(\mathit{ini} \rightarrow \Diamond \mathit{fin})$,
where $\mathit{ini}$ and $\mathit{fin}$ denote start and end
(\texttt{SZS = Theorem}).
\emph{Safety} used $\Box \neg(A \wedge B)$ to forbid co-occurrence of activities
(e.g., $A=$ ``accept'', $B=$ ``reject''), also with \texttt{SZS = Theorem}.
All three were tested independently on the same coalition, yielding three outcomes $v(C)$: sat, liv, saf.

\begin{figure}[!htb]
  \centering
  \includegraphics[width=1\linewidth]{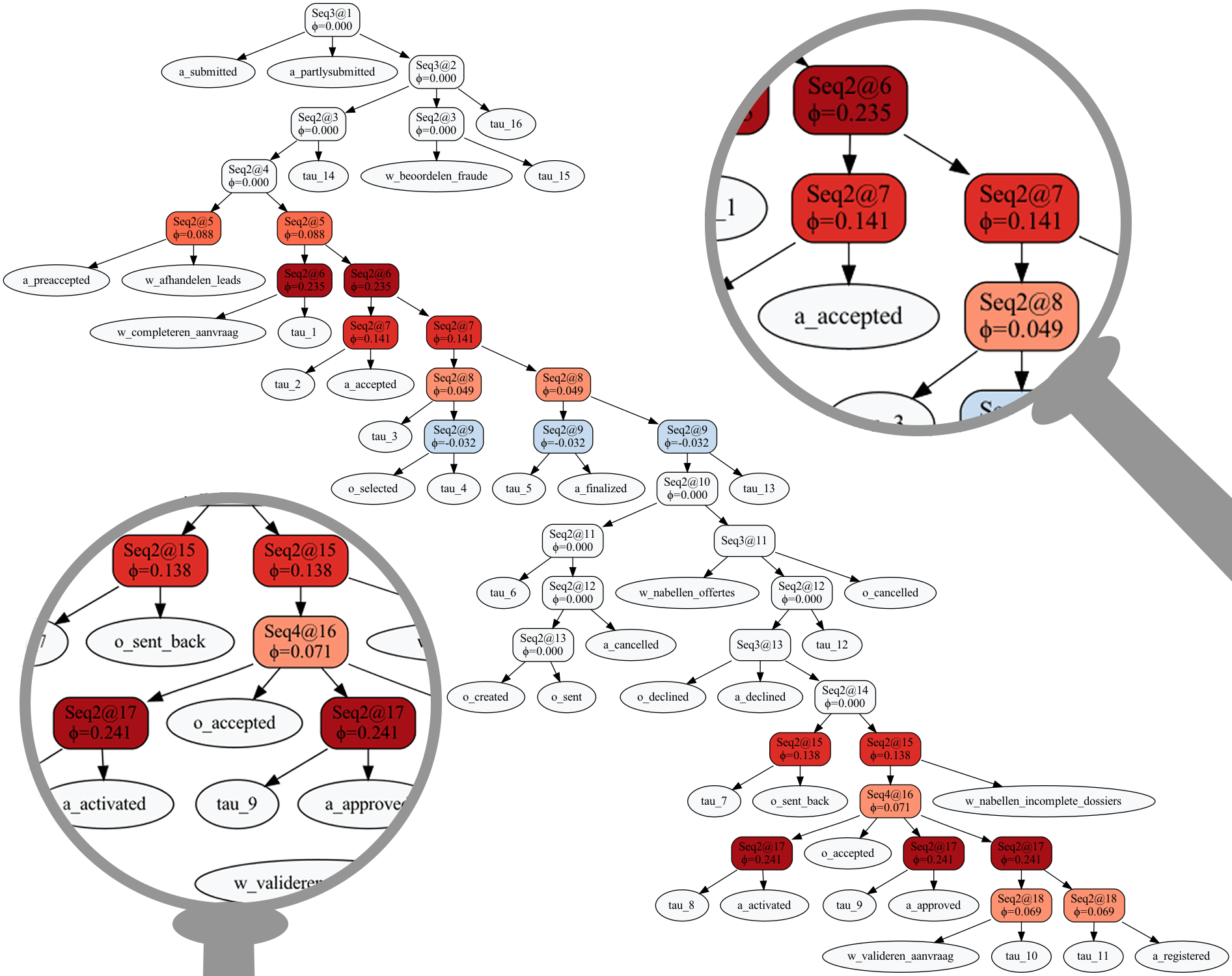}
\caption{Workflow tree mined from the BPI dataset at noise level~0.5,
with nodes colored by Shapley values indicating their relative contribution to logical properties.
Red nodes are critical, blue nodes redundant or harmful.}
\label{fig:shapley-tree}
\end{figure}

Figure~\ref{fig:shapley-tree} shows how attribution scores
translate into an intuitive visualization: nodes with strong positive
impact appear in red, while negligible or negative ones fade toward blue.
The tree reveals at a glance which patterns are indispensable,
which can be ignored, and which undermine property outcomes,
providing a diagnostic view of workflow quality.

\subsection{Findings}

Table~\ref{tab:top3} reports the \emph{Top-3 elements} per dataset and property,
ranked by average $|\phi|$ across noise levels (MC approximation).
\begin{table}[!htb]
  \centering
  \caption{Top-3 elements per dataset and property (ranked by average $|\phi|$; MC approximation).}
  \label{tab:top3}
  \hspace*{-3mm}
  \begin{tabular}{lp{0.29\linewidth}p{0.337\linewidth}}
    \toprule
    Dataset / Property & Top-3 Elements & Interpretation \\
    \midrule
    BPI\,2012 (satisfiability) & Seq2@18, Seq2@16, Seq2@19 & Two dominant nodes drive satisfaction. \\
    Hospital (satisfiability) & Seq2@6, Seq2@2, Seq2@13 & Clear critical nodes; others marginal. \\
    Hospital (liveness) & Seq2@13, Seq2@15, Seq2@6 & Lower scores; progress less sensitive. \\
    Running Example (all) & All $\approx 0$ & Toy model; no critical nodes. \\
    \bottomrule
  \end{tabular}
\end{table}


Table~\ref{tab:mc-vs-rs} summarizes the trade-offs between Monte Carlo (MC)
and Random Subset (RS) sampling. MC closely follows the theoretical definition but is computationally heavy,
while RS is much cheaper but introduces bias and variance.

\begin{table}[!htb]
\centering
\caption{Comparison of Monte Carlo (MC) and Random Subset (RS) sampling.}
\label{tab:mc-vs-rs}
\begin{tabular}{p{1.6cm}p{3.05cm}p{3.05cm}}
\toprule
 & \textbf{Monte Carlo (MC)} & \textbf{Random Subset (RS)} \\
\midrule
\textbf{Stability} & $\Delta_{\text{max}}<0.01$ after 1{,}000 permutations; smooth convergence & Higher variance; $\Delta_{\text{max}}$ may oscillate \\
\textbf{Bias} & Unbiased; exact Shapley semantics & Biased; favors large coalitions \\
\textbf{Efficiency} & Many prover queries; expensive & Far fewer queries; cheap \\
\textbf{Use case} & Accuracy and stability critical & Fast, approximate results \\
\bottomrule
\end{tabular}
\end{table}

The experiments reveal several observations:
\begin{itemize}
  \item \textbf{Critical vs.\ redundant elements.} Few nodes consistently emerge as critical; most contribute negligibly.
  \item \textbf{Harmful elements.} Negative Shapley values identify nodes that reduce property satisfaction.
  \item \textbf{Effect of noise.} In BPI\,2012 and Hospital Billing, noise amplifies dispersion and raises average values; the Running Example stays flat.
  \item \textbf{Property-specific sensitivity.} Satisfiability correlates strongly with noise, liveness varies, safety remains robust.
  \item \textbf{Ranking stability.} Top-5 rankings are stable across noise levels and approximations, with Jaccard similarity~\cite{Oh-etal-2022} above 0.8.
  \item \textbf{Approximation trade-offs.} MC yields more stable estimates, RS is an order of magnitude faster with similar Top-$k$ results.
\end{itemize}

\begin{figure}[!htb]
  \centering
  \includegraphics[width=.8\linewidth]{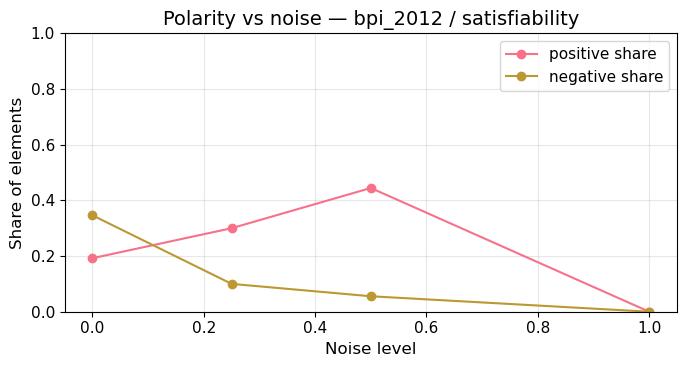}
  \caption{Effect of noise on positive vs.\ negative Shapley values.}
  \label{fig:pos-neg}
\end{figure}


\begin{figure*}[!htb]
  \centering
  \includegraphics[width=\textwidth]{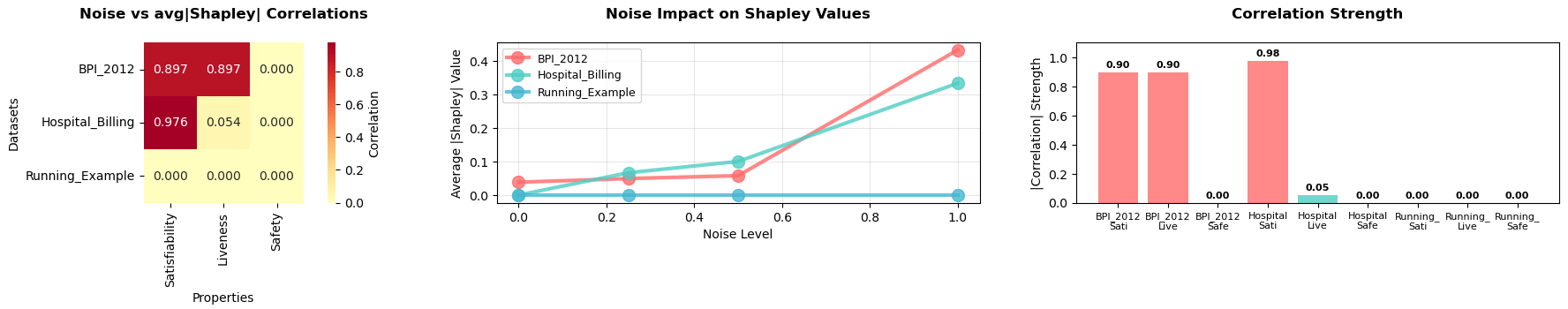}
  \caption{Impact of noise on Shapley values across datasets.
  Left: correlation between noise and average $|\phi|$.
  Middle: average Shapley values vs.\ noise.
  Right: correlation strength by dataset and property.}
  \label{fig:noise-impact}
\end{figure*}


Figure~\ref{fig:noise-impact} shows that satisfiability and liveness in
BPI\,2012 and Hospital Billing correlate strongly with noise
($\text{corr}=0.897$ and $0.976$), while safety remains unaffected.
Increasing noise raises average Shapley values sharply
(e.g., BPI\,2012 from $0.039$ at noise $0.0$ to $0.432$ at $1.0$),
making more nodes appear important but reducing interpretability.
The Running Example remains flat. Near-zero correlations, as for safety, indicate robustness:
rankings remain stable regardless of log quality.
For analysts, this distinction is critical — high-correlation models demand cleaner logs,
while low-correlation ones remain reliable under noise.

\subsection{Interpretation of Results}

This study outlines an emerging direction toward explainable workflow verification,
where outcomes are not only computed but also interpreted.

Our experiments show that Shapley-based attribution can distinguish
\emph{critical}, \emph{redundant}, and even \emph{harmful} elements in mined workflows,
adding a qualitative layer beyond binary property checks.
We also observe dataset-specific behavior: in BPI\,2012, noise amplifies the dispersion of importance;
in Hospital Billing, it simplifies the model; and in the Running Example, it has little effect.

These results, though preliminary, suggest that Shapley-based analysis can support diagnosis,
guide simplification, and open a path toward tool-supported explainable workflow verification.

\begin{table*}[!htb]
\centering
\caption{Model quality analysis under three perspectives.}
\label{tab:quality}
\begin{tabular}{lcccccccc}
\toprule
\textbf{Analysis Type} &
\textbf{Critical Th.} &
\textbf{Redundant Th.} &
\textbf{Critical Elem.} &
\textbf{Critical \%} &
\textbf{Neutral Elem.} &
\textbf{Neutral \%} &
\textbf{Redundant Elem.} &
\textbf{Redundant \%} \\
\midrule
Standard        & $\geq 0.1$  & $\leq 0.01$  & 28 & 6.6 & 28 & 6.6 & 367 & 86.8 \\
High Sensitivity& $\geq 0.05$ & $\leq 0.005$ & 36 & 8.5 & 35 & 8.3 & 352 & 83.2 \\
Low Sensitivity & $\geq 0.2$  & $\leq 0.02$  & 20 & 4.7 & 23 & 5.4 & 380 & 89.8 \\
\bottomrule
\end{tabular}
\end{table*}

\subsection{Model Diagnostics}

Beyond binary checks, our framework offers diagnostic views that make workflows more interpretable.
We track which control-flow patterns (Seq, And, Xor, Loop) and which properties (satisfiability, liveness, safety) dominate attribution scores,
and how they shift with noise, revealing which constructs \emph{drive} correctness.
Nodes whose Shapley values change sign or magnitude under noise are labeled \emph{adaptive}, highlighting sensitivity to data quality and distinguishing genuine structures from discovery artifacts.
A trusted baseline (e.g., noise=0.0) serves as a reference to detect anomalies and assess stability.

Using thresholds (e.g., $|\phi|\geq 0.05$), we classify nodes as \emph{critical}, neutral, or redundant, enabling systematic partitioning of models.
Finally, we propose a three-perspective diagnosis (standard, high sensitivity, low sensitivity), which varies thresholds to highlight robustness, risk points, or redundancy reduction.
Table~\ref{tab:quality} shows that across benchmarks, only 5--9\% of elements are critical, while over 85\% are redundant.

Overall, these diagnostics extend explainable verification with practical tools:
they identify patterns that drive correctness, expose fragile or harmful nodes, and provide perspectives on model quality that analysts can adapt to their needs.

\subsection{Limitations}

This work is exploratory and has several limitations.
Exact Shapley values are infeasible for large workflows, so we rely on approximations (Monte Carlo and random subsets) with inherent trade-offs between stability and efficiency.
The focus on satisfiability, liveness, and safety covers only part of the relevant property space, leaving richer temporal and quantitative conditions for future work.
Finally, the evaluation is limited to benchmark datasets; broader validation on industrial logs is needed to assess generalizability.

\subsection{Potential Applications}

Despite these limitations, the approach has several practical applications:
\begin{itemize}
  \item \textbf{Model diagnosis:} identifying nodes essential for correctness or harmful to property outcomes.
  \item \textbf{Simplification:} removing redundant structures while preserving key properties.
  \item \textbf{Explainability:} making property outcomes transparent to stakeholders.
  \item \textbf{Process optimization:} guiding improvements by showing which elements drive satisfiability, liveness, or safety.
  \item \textbf{Compliance:} highlighting components critical for regulatory requirements.
\end{itemize}

\subsection{Future Work}

We see this as a foundation for bridging formal methods and explainable AI in process mining. Next steps include:
\begin{itemize}
  \item scaling approximation strategies to industry-scale workflows,
  \item extending properties toward fairness, quantitative constraints, and compliance rules,
  \item benchmarking against other explainability techniques from ML,
  \item validating on industrial logs in software analytics and business process management,
  \item developing interactive tools that let engineers explore explanations, run what-if analyses, and optimize workflows collaboratively.
\end{itemize}

\section{Conclusion}

This work outlines an emerging direction at the intersection of
formal methods and explainability.
By combining logical property analysis with Shapley-based attribution,
we move beyond binary correctness answers toward interpretable rankings of workflow elements.
This opens a path to bridging formal verification and explainable AI for process mining,
with direct relevance for software engineering through compliance checks,
redundancy reduction, process optimization, and stakeholder-facing explanations.

\bibliography{bib-rk,bib-rk-main,bib-rk-tools}

@ARTICLE{Alpern-Schneider-1985,
  author = {Alpern, Bowen and Schneider, Fred B.},
  title = {Defining liveness},
  journal = {Information Processing Letters},
  year = {1985},
  volume = {21 (4)},
  pages = {181--185}
}

@book{Manna-Pnueli-1992,
 author = {Manna, Zohar and Pnueli, Amir},
 title = {The Temporal Logic of Reactive and Concurrent Systems -- Specification},
 year = {1992},
 publisher = {Springer-Verlag New York, Inc.},
}

@article{Kindler-1994,
    Author = {Kindler, Ekkart},
    Title = {Safety and Liveness Properties: A Survey},
    Journal = {EATCS-Bulletin},
    Volume = {53},
    Year = {1994},
    Pages = {}
}

@misc{Schreiner-2023,
    author = {Schreiner, Wolfgang},
    title = {First-order logic: software for proving, Course 'Computational logic',
            \url{https://moodle.risc.jku.at/pluginfile.php/11902/mod_resource/content/9/10-fol6.pdf}. The last access 25.04.2024},
    year = {2023},
}

@Article{Sutcliffe-2017,
   Author =	 "Sutcliffe, Geoff",
   Title =	 "{The TPTP Problem Library and Associated Infrastructure}",
   Journal =	 "Journal of Automated Reasoning",
   Year =	 2017,
   Volume =	 59,
   Pages =	 "438--502",
}

@inproceedings{Leemans-etal-2013,
  author       = {Sander J. J. Leemans and
                  Dirk Fahland and
                  Wil M. P. van der Aalst},
  editor       = {Niels Lohmann and
                  Minseok Song and
                  Petia Wohed},
  title        = {Discovering Block-Structured Process Models from Event Logs Containing
                  Infrequent Behaviour},
  booktitle    = {Business Process Management Workshops - {BPM} 2013 International Workshops,
                  Beijing, China, August 26, 2013, Revised Papers},
  series       = {Lecture Notes in Business Information Processing},
  volume       = {171},
  pages        = {66--78},
  publisher    = {Springer},
  year         = {2013},
  doi          = {10.1007/978-3-319-06257-0\_6},
}

@article{Riazanov-Voronkov-2002,
author = {Riazanov, Alexandre and Voronkov, Andrei},
title = {The design and implementation of {VAMPIRE}},
year = {2002},
publisher = {IOS Press},
address = {NLD},
volume = {15},
number = {2,3},
issn = {0921-7126},
journal = {Journal of AI Communications},
pages = {91--110},
numpages = {20},
}

@article{Schulz-2002,
author = {Schulz, Stephan},
title = {E -- a brainiac theorem prover},
year = {2002},
issue_date = {August 2002},
publisher = {IOS Press},
volume = {15},
number = {2,3},
issn = {0921-7126},
journal = {Journal of AI Communications},
month = {aug},
pages = {111–126},
numpages = {16},
}

@article{Augusto-etal-2019,
author = {Augusto, Adriano and Conforti, Raffaele and Dumas, Marlon and Rosa, Marcello La and Maggi, Fabrizio Maria and Marrella, Andrea and Mecella, Massimo and Soo, Allar},
title = {Automated Discovery of Process Models from Event Logs: Review and Benchmark},
year = {2019},
issue_date = {April 2019},
journal = {IEEE Transactions on Knowledge and Data Engineering},
publisher = {IEEE Educational Activities Department},
address = {USA},
volume = {31},
number = {4},
doi = {10.1109/TKDE.2018.2841877},
month = {apr},
pages = {686–705},
numpages = {20}
}

@inproceedings{Oh-etal-2022, 
  series={CIKM ’22},
   title={Rank List Sensitivity of Recommender Systems to Interaction Perturbations},
   url={http://dx.doi.org/10.1145/3511808.3557425},
   DOI={10.1145/3511808.3557425},
   booktitle={Proceedings of the 31st ACM International Conference on Information and Knowledge Management},
   publisher={ACM},
   author={Oh, Sejoon and Ustun, Berk and McAuley, Julian and Kumar, Srijan},
   year={2022},
   month=oct, pages={1584–1594},
   collection={CIKM ’22} 
}

@inproceedings{Lundberg-Lee-2017,
author = {Lundberg, Scott M. and Lee, Su-In},
title = {A unified approach to interpreting model predictions},
year = {2017},
isbn = {9781510860964},
publisher = {Curran Associates Inc.},
address = {Red Hook, NY, USA},
booktitle = {Proceedings of the 31st International Conference on Neural Information Processing Systems},
pages = {4768--4777},
numpages = {10},
location = {Long Beach, California, USA},
series = {NIPS'17}
}

@article{Rozinat-vanderAalst-2008,
  author    = {Anne Rozinat and Wil M. P. van der Aalst},
  title     = {Conformance Checking of Processes Based on Monitoring Real Behavior},
  journal   = {Information Systems},
  volume    = {33},
  number    = {1},
  pages     = {64--95},
  year      = {2008},
  doi       = {10.1016/j.is.2007.07.001}
}

@book{van-der-Aalst-2016-ProcessMining,
  author    = {Wil M. P. van der Aalst},
  title     = {Process Mining: Data Science in Action},
  publisher = {Springer},
  edition   = {2nd},
  year      = {2016},
  doi       = {10.1007/978-3-662-49851-4}
}

@inproceedings{Ribeiro-etal-2016-LIME,
  author    = {Marco Tulio Ribeiro and Sameer Singh and Carlos Guestrin},
  title     = {"Why Should I Trust You?": Explaining the Predictions of Any Classifier},
  booktitle = {Proceedings of the 22nd ACM SIGKDD International Conference on Knowledge Discovery and Data Mining (KDD)},
  pages     = {1135--1144},
  year      = {2016},
  doi       = {10.1145/2939672.2939778}
}

@misc{Vampire-tool,
    author = {Andrei Voronkov},
    title = {Website for prover {Vampire}},
     url={https://vprover.github.io/},
    year = {2017},
    note =         {accessed on 5-Aug-2024},
}

@misc{E-tool,
    author = {Stephan Schulz},
    title = {Website for prover {E}},
     url={http://wwwlehre.dhbw-stuttgart.de/\textasciitilde sschulz/E/E.html},
    year = {2020},
    note =         {accessed on 5-Aug-2024},
}

@inproceedings{Klimek-Witek-2024-ASE-RENE,
author = {Klimek, Radoslaw and Witek, Julia},
title = {Automatic Generation of Logical Specifications for Behavioural Models},
year = {2024},
isbn = {9798400712708},
publisher = {Association for Computing Machinery},
address = {New York, NY, USA},
url = {https://doi.org/10.1145/3695750.3695822},
doi = {10.1145/3695750.3695822},
booktitle = {Proceedings of the 39th IEEE/ACM International Conference on Automated Software Engineering Workshops (ASE/RENE), Sun 27 October--Fri 1 November 2024, Sacramento, CA, USA},
pages = {1--7},
numpages = {7},
location = {Sacramento, CA, USA},
series = {ASEW'24}
}

\end{document}